\documentstyle{article}

\def\be{\begin{equation}}
\def\ee{\end{equation}}
\def\bea{\begin{eqnarray}}
\def\eea{\end{eqnarray}}
\begin{document}

\pagestyle{empty}
\vskip-10pt
\hfill G\"{o}teborg ITP 99-07
\vskip-10pt
\hfill {\tt hep-th/9906053}
\begin{center}
\vskip 3truecm
{\Large\bf
The light spectrum near the Argyres-Douglas point
}\\ 
\vskip 2truecm
{\large\bf
Andreas Gustavsson and M{\aa}ns Henningson
}\\
\vskip 1truecm
{\it Institute of Theoretical  Physics,
Chalmers University of Technology, \\
S-412 96 G\"{o}teborg, Sweden}\\
\vskip 5truemm
{\tt f93angu@fy.chalmers.se, mans@fy.chalmers.se}
\end{center}
\vskip 2truecm
\noindent{\bf Abstract:}
We consider $N = 2$ super Yang-Mills theory with $SU(2)$ gauge group and a single quark hypermultiplet in the fundamental representation. For a specific value of the quark bare mass and at a certain point in the moduli space of vacua, the central charges corresponding to two mutually non-local electro-magnetic charges vanish simultaneously, indicating the possibility of massless such states in the spectrum. By realizing the theory as an $M$-theory configuration, we show that these states indeed exist in the spectrum near the critical point.    
\vfill
\vskip4pt
\noindent{June 1999}

\eject
\newpage
\pagestyle{plain}

\section{Introduction}
During the last few years, the low-energy effective theories of $N = 2$ supersymmetric Yang-Mills theories in $d = 4$ space-time dimensions with various gauge groups and matter representations have been subject of much interest, following \cite{SW}. At a generic point on the Coulomb branch of the moduli space of vacua, the massless degrees of freedom constitute $r$ free $U(1)$ vector multiplets, where $r$ is the rank of the microscopic gauge group. The dynamics of these multiplets is encoded in the special K\"ahler geometry of the Coulomb branch. The states of the theory are characterized by their electric and magnetic charges $n_e$ and $n_m$ with respect to the $U(1)$ gauge fields and possibly also some quark number charges $S$ (if the microscopic theory contains hyper multiplet matter). Given two such states with quantum numbers $(n_e, n_m, S)$ and $(n_e', n_m', S')$ respectively, we can form the `symplectic' product
\be
c = n_e \cdot n_m^\prime - n_m \cdot n_e^\prime .
\ee
To see the physical significance of $c$, we can consider quantizing one of the particles in the presence of the other. The wave `function' is then in fact a section of a non-trivial line bundle over (punctured) space, the first Chern class of which, when evaluated on an $S^2$, equals $c$. We say that the states are mutually non-local if $c$ is non-zero.  

The mass $M$ of an arbitrary state obeys the inequality
\be
M \geq |Z| , \label{BPS}
\ee
where $Z$ is the the central charge that appears in the $N = 2$ supertranslations algebra. The central charge is in general a linear combination of conserved Abelian charges, e.g.
\be
Z = a \cdot n_e + a_D \cdot n_m + m \cdot S,
\ee
where the coefficients $a$, $a_D$ and $m$ are some holomorphic functions of the moduli and parameters of the theory. States which saturate the bound (\ref{BPS}), so called BPS states, are of particular interest. Being the lightest states in their charge sector, generically they cannot disappear as the moduli and parameters of the theory are varied. (The exception is when a domain wall of `marginal stability' is reached, where the phases of the central charges of three BPS states are equal. It might then be possible for the heaviest particle to decay into the two lighter ones and be absent from the spectrum on the other side of the domain wall.) 

Along certain submanifolds of the Coulomb branch, the central charges corresponding to certain sets of quantum numbers vanish. The corresponding BPS states, if they are present in the spectrum of the theory, would then be massless. The nature of the low energy effective theory depends crucially on the electric and magnetic quantum numbers of these states. If the symplectic product $c$ vanishes for any pair of massless particles, it is always possible to perform an electric/magnetic duality transformation after which all massless states have purely electric charges. The low energy effective theory would then be massless $N = 2$ supersymmetric QED with some number of different flavours of matter. Our focus in this paper is the case when at least one symplectic product does not vanish, so that the theory necessarily contains both electrically and magnetically charged massless states. The low energy effective theory is then an exotic interacting $N = 2$ superconformal theory with no known Lagrangian description. 

An example where such a phenomenon could possibly occur was first discovered in pure $SU(3)$ super Yang-Mills theory, where the central charges for mutually non-local states vanishes at a certain critical point in the moduli space \cite{AD}. An equivalent critical point can be found in the moduli space of $SU(2)$ super Yang-Mills theory with a single hyper multiplet in the fundamental representation \cite{APSW}. There are also more complicated examples known. In all these examples, the relevant BPS states are known to exist in the spectrum in the weak-coupling region, where semi-classical methods can be trusted. The critical point is at strong coupling, though, and it is conceivable that these states have decayed at some domain wall of marginal stability before this point is reached. The purpose of this paper is to investigate a theory close to such a critical point to see whether the spectrum really contains arbitrarily light mutually non-local states.

Given a set of quantum numbers $(n_e, n_m, S)$, it is in general a difficult problem to determine if a corresponding BPS state exists in the spectrum (at a given point in the moduli space) \cite{Bilal}. In principle, this question can be answered by realizing the theory as an $M$-theory configuration \cite{Witten97}. We thus consider $M$-theory on an eleven-manifold $M^{1,10}$ of the form
\be
M^{1, 10} \simeq {\bf R}^{1, 3} \times {\bf R}^3 \times Q^4 ,
\ee
where the first factor is four-dimensional Minkowski space, the second factor is three dimensional Euclidean space, and the last factor is some four-dimensional manifold of $SU(2)$ holonomy. We also introduce an $M$-theory five-brane with a world-volume of the form $W^{1, 5}$ of the form
\be
W^{1, 5} \simeq {\bf R}^{1, 3} \times p \times \Sigma ,
\ee
where $p$ is a point in ${\bf R}^3$ and $\Sigma$ is a two-manifold in $Q^4$. The theory on $W^{1, 5}$ then has an effective  four-dimensional low-energy limit on ${\bf R}^{1, 3}$, and this is the theory we are interested in. Excitations around the vacuum defined by $M^{1, 10}$ and $W^{1, 5}$ are described by $M$-theory two-branes (membranes) with world-volumes $S^{1, 2}$ of the form
\be
S^{1, 2} \simeq \Gamma \times p \times D ,
\ee
where $\Gamma$ defines the world-line of a particle in ${\bf R}^{1, 3}$ and $D$ is a two-manifold in $Q^4$ whose boundary $C = \partial D$ lies on $\Sigma$. (That a two-brane can end on a five-brane has been shown in \cite{Strominger}.) The mass $M$ of the state is given by
\be
M = 2 \int_D V_D ,
\ee
where $V_D$ is the volume-form of $D$. The homology class $[C]$ of $C$ determines the quantum numbers $(n_e, n_m, S)$ of the state. In particular, the intersection number $[C] \cdot [C^\prime]$ of two homology classes equals the symplectic product of the corresponding electric and magnetic charges, i.e.
\be
[C] \cdot [C^\prime] = n_e \cdot n_m^\prime - n_m \cdot n_e^\prime .
\ee  

The $SU(2)$ holonomy of $Q^4$ means that this space is hyper K\"ahler, i.e. it admits a two-sphere $S^2$ of inequivalent complex structures $J$. Equivalently, it can be regarded as a Ricci-flat K\"ahler manifold, and therefore admits a covariantly constant holomorphic two-form $\Omega$. The relationship between these two descriptions is as follows: Given a complex structure $J$, we have $\Omega \sim K^\prime +i K^{\prime \prime}$, where $K^\prime$ and $K^{\prime \prime}$ are the K\"ahler forms corresponding to two other complex structures $J^\prime$ and $J^{\prime \prime}$ such that $J$, $J^\prime$ and $J^{\prime \prime}$ are all orthogonal. The requirement that the effective theory on ${\bf R}^{1, 3}$ has $N = 2$ supersymmetry is equivalent to demanding that $\Sigma$ be holomorphically embedded in $Q^4$ with respect to some complex structure $J$ \cite{Henningson-Yi}. The central charge $Z$ of a state corresponding to a two-manifold $D$ is then given by
\be
Z = \int_D \Omega_D ,
\ee
where $\Omega_D$ is the pullback (by the embedding map) of $\Omega$ to $D$. One can show that the inequality (\ref{BPS}) holds. It is saturated, i.e. the state is BPS, if $D$ is holomorphically embedded with respect to a complex structure $J^\prime$ which is orthogonal to $J$. Given $J$, there is a circle $S^1$ of such $J^\prime$, corresponding to the phase of the central charge $Z$.

Before we answer the question of the light spectrum near a critical point, we must decide exactly what states we are looking for. The strongest result would be to establish the existence of mutually non-local BPS states, the masses of which of course vanish as we approach the critical point where their central charges vanish. This is fairly difficult, though, since one would have to construct the corresponding two-manifolds $D$ exactly. The requirements that $D$ be holomorphically embedded with respect to the appropriate complex structure and also intersects $\Sigma$ along a real curve are difficult to analyze, except in certain special situations. An alternative would be to simply find mutually non-local states, whose masses vanish at the critical point but which are not BPS. This would be a fairly weak result: Given two such states, one could construct a third such state by simply taking the sum of the corresponding two-manifolds $D$. We will instead consider an intermediate set of states that we call asymptotically BPS. By this we mean that 
\be
M / |Z| \rightarrow 1
\ee
as we approach the critical point. This implies that $M$ goes to zero as we approach the critical point. However, the sum of two such states is in general not of the same kind. It seems plausible that the existence of an asymptotically BPS state implies the existence of a BPS state, but we have no proof of this. 

\section{The computation}
We will consider the case of $SU(2)$ super Yang-Mills theory with one hypermultiplet quark in the fundamental representation. As explained in \cite{Witten97}, the hyper K\"ahler manifold $Q^4$ is in this case simply ${\bf R}^3 \times S^1$ with coordinates $X^4$, $X^5$, $X^6$ and $X^{10}$, where $X^{10}$ is periodic with period $2 \pi$. The complex structure $J$ can be described by declaring that $s = X^6 + i X^{10}$ and $v = X^4 + i X^5$ are holomorphic coordinates. The K\"ahler form $K$ and the covariantly constant holomorphic two-form $\Omega$ are then given by
\bea
K & = & i \left( d s \wedge d \bar{s} + d v \wedge d \bar{v} \right) \cr
\Omega & = & 2 d s \wedge d v .
\eea
It is convenient to replace $s$ by the single-valued coordinate $t = \exp (-s)$. Then the complex surface $\Sigma$ can be written as (\cite{Hanany}, \cite{APS})
\be
t^2 + (v^2 - \phi^2) t + \Lambda^3 (v - m) = 0 .
\ee
Here $m$ is the bare mass of the quark hypermultiplet, and the modulus $\phi^2$ parametrizes the moduli space of vacua. $\Lambda$ is the dynamically generated scale of the theory, which we will choose so that $\Lambda^3 = -2$. If we solve the equation for $s = - \log t$ we then get
\be
s = - \log \left(-\frac{1}{2} (v^2 - \phi^2) + \sqrt{f} \right) ,
\ee
where
\be
f = \frac{1}{4} (v^2 - \phi^2)^2 + 2 (v - m) .
\ee
We see that $\Sigma$ can be thought of as a double cover of the $v$-plane (because of the ambiguity of the square root). The two sheets meet at the four branch points, joined by branch cuts, where $f$ vanishes. 

Generically, the branch points lie at different values of $v$, but by choosing $m = \frac{3}{2}$ and $\phi^2 = 3$ (the Argyres-Douglas point), we get $f = \frac{1}{4} (v - 1)^3 (v + 3)$, i.e. three of the branch points coincide at $v = 1$. We are interested in the geometry close to this point, so we write
\bea
m & = & \frac{3}{2} + \delta m \cr
\phi^2 & = & 3 + \delta \phi^2 \cr
v & = & 1 + u .
\eea
Dropping all non-leading terms in $\delta m$, $\delta \phi^2$ and $u$, we then get
\be
s = u - \sqrt{f} ,
\ee
where
\be
f = u^3 - (\delta v)^3 
\ee
and $\delta v$ is defined so that
\be
(\delta v)^3 = 2 \delta m - \delta \phi^2 .
\ee
The three branch points are thus located at $u = \omega \delta v$, where $\omega$ is an arbitrary cubic root of unity. We define the homology classes $[C_1]$, $[C_2]$, and $[C_3]$ by representative curves $C_1$, $C_2$, $C_3$ that winds tightly around two of these branch points but not the third. (Only two of these classes are linearly independent, though.) It is easy to see that the intersection form of these curves is given by
\be
\left( \begin{array}{ccc}0 & 1 & -1\\-1 & 0 & 1\\1 & -1 & 0 \end{array} \right)\label{intersection}
\ee
so that any two of the corresponding states are mutually non-local. 
 
To construct the surfaces corresponding to BPS states it is convenient to change variables from $s$ and $u$ to $x$ and $y$ defined by the relations
\bea
s & = & (x + y) \delta v \cr
u & = & (x - y) \delta v .
\eea
The equation for $\Sigma$ then reads
\be
(x + y) \delta v = (x - y) \delta v - \sqrt{(x- y)^3 - 1} (\delta v)^{3/2} + {\cal O}((\delta v)^2) .
\ee
Solving for $y$, we get
\be
y = - \frac{1}{2} \sqrt{x^3 - 1} (\delta v)^{1/2} + {\cal O}(\delta v) .
\ee
In these variables, the K\"ahler form and the holomorphic two-form are
\bea
K & = & 2 i |\delta v|^2 \left(dx \wedge d \bar{x} + d y \wedge d \bar{y} \right) \cr
\Omega & = & 4 (\delta v)^2 d y \wedge d x .
\eea

We now consider a surface $D$ with a boundary $C$ on $\Sigma$ that winds around the two branch points at $x = \omega$ and $x = \omega^\prime$, where $\omega$ and $\omega^\prime$ are two different cubic roots of unity, but not the third. The central charge is independent of the exact form of $D$, so we can parametrize it by the real parameters $\xi$ and $\eta$ such that $-1 \leq \xi \leq 1$ and $-1 \leq \eta \leq 1$ and let it be of the form
\bea\label{xyAnsatz}
x & = & x (\xi) \cr
y & = & \frac{\eta}{2} \sqrt{x^3 (\xi) - 1} (\delta v)^{1/2} ,
\eea
where the function $x (\xi)$ obeys $x (-1) = \omega$ and $x (1) = \omega^\prime$. The central charge is then given by
\be\label{ZD}
Z = 4 (\delta v)^2 \int_\omega^{\omega^\prime} d x \int_{-1}^1 \frac{d \eta}{2} \sqrt{x^3 - 1} (\delta v)^{1/2} = 4 (\delta v)^{5/2} \int_\omega^{\omega^\prime} d x \sqrt{x^3 - 1} .
\ee
In particular, $|Z| \rightarrow 0$ as $\delta v \rightarrow 0$. Notice that $\Sigma$ in the limit $\delta v \rightarrow 0$ has $Z_2 X Z_3$ symmetry: $y \rightarrow -y$ and $x \rightarrow \exp (2\pi i/3) x$, and is a flat sheet: $y=0$. The three surfaces $D$ must break the $Z_3$ but have the $Z_2$ symmetry, and they must intersect with $\Sigma$ orthogonally. The surface (\ref{xyAnsatz}) has these properties, and it will turn out that the surface that is asymptotically BPS is really of this form. The induced metric on such a surface is 
\bea
d s^2 & = & 2 |\delta v|^2 \left(|d x|^2 + |d y|^2 \right) \cr
& = & g_{\xi \xi} d \xi^2 + 2 g_{\xi \eta} d \xi d \eta + g_{\eta \eta} d \eta^2 ,
\eea
 where the components are
\bea
g_{\xi \xi} & = & 2 |\delta v|^2 |x^\prime (\xi)|^2 + {\cal O} ((\delta v)^3) \cr
g_{\xi \eta} & = & {\cal O} ((\delta v)^3) \cr
g_{\eta \eta} & = & \frac{|\delta v|^3}{2} |\sqrt{x^3 (\xi) - 1}|^2 + {\cal O} ((\delta v)^4) .
\eea 
The mass is then
\be\label{AD}
M = 2 \int_{-1}^1 d \eta \int_{-1}^1 d \xi \sqrt{\det g} = 4 |\delta v|^{5/2} \int_{-1}^1 d \xi |x^\prime (\xi) \sqrt{x^3 (\xi) - 1}| + {\cal O} ((\delta v)^3) .
\ee

Comparing (\ref{ZD}) and (\ref{AD}), we see that $M / |Z| \rightarrow 1$ as $\delta v \rightarrow 0$ provided that the function $x (\xi)$ is chosen so that the phase of
\be\label{integrand}
x^\prime (\xi) \sqrt{x^3 (\xi) - 1}
\ee
is independent of $\xi$. We have not succeeded in finding such functions analytically, but they are not difficult to approximate numerically. For example, if we take $\omega = \exp (-2 \pi i / 3)$ and $\omega^\prime = \exp 2 \pi i / 3$, we find that the central charge (\ref{ZD}) is real, i.e. symmetric under complex conjugation. The same should be true for the expression (\ref{integrand}), from which follows that ${\rm Re} \, x (\xi)$ is an even and ${\rm Im} \, x (\xi)$ an odd function of $\xi$. We can then expand $x (\xi)$ around $\xi = -1$ as
\be
x (\xi) = \exp (- 2 \pi i / 3) + x^\prime (-1) (\xi + 1) + {\cal O} ((\xi + 1)^2) ,
\ee
where the phase of $x^\prime (-1)$ equals $4 \pi / 9$. The curve $x (\xi)$ can be well approximated by a rather flat parabola. Rotating the configuration through $2 \pi / 3$ around the origin of the $x$-plane, we obtain the solutions for the two other homology classes. Hence we have found the three mutually non-local asymptotic BPS states. 
  
\vskip 0.5truecm
We have benefited from discussions with Philip Argyres and Piljin Yi. The research of M. H. is supported by the Swedish Natural Science Research Council (NFR).

\end{document}